\documentstyle[prd,aps,preprint]{revtex}
\tightenlines
\newcommand {\be}{\begin{eqnarray}}
\newcommand {\ee}{\end{eqnarray}}
\input epsf.tex
\def\DESepsf(#1 width #2){\epsfxsize=#2 \epsfbox{#1}}
\begin{document}

\draft
\preprint{\vbox{
\hbox{UMD-PP-00-066}
}}
\title{Mirror Dark Matter}
\author{ R. N. Mohapatra$^1$ and V. L. Teplitz$^{2,3}$ }

\address{$^1$Department of
Physics, University of Maryland, College Park, MD, 20742\\
$^2$ Office of Science and Technology, Old Executive Office Building,
Washington DC, 20502\\
$^3$ Department of Physics, Southern Methodist University, Dallas,
TX-75275 (permanent address).}
\date{March, 2000}
\maketitle
\begin{abstract}
There appear to be three challenges that any theory of dark matter must
face: (i) why is $\Omega_{DM}$ of the same order as $\Omega_{Baryons}$ ?
(ii) what are the near solar mass objects ($\sim 0.5 M_{\odot}$) observed
by the MACHO microlensing project ? and (iii) understanding the shallow
core density profile of the halos of dwarf as well as low surface
brightness galaxies. The popular cold dark matter candidates,
the SUSY LSP and the axion fail to meet these challenges. We argue that
in the mirror model suggested recently to explain the neutrino anomalies,
the mirror baryons being 15-20 times heavier than familiar baryons, can 
play the role of the cold dark matter and provide reasonable
explanation of all three above properties without extra assumptions.
\end{abstract}

\section{Introduction}

Dark matter constitutes the bulk of the matter in the universe and a
proper understanding of the nature of the new particle that plays this
role has profound implications not only for cosmology but also for
particle physics beyond the standard model\cite{book}. It is therefore not
surprising that one of the major areas of research in both particle
physics and cosmology continues to be the physics of dark
matter.

Apart from the simple requirement that the right particle physics
candidate must have properties that allow it to have the requisite relic
density and mass to dominate the mass content of the universe, it
should be required to
 provide a satisfactory resolution of three puzzles of dark matter
physics: (i) why is it that the contribution of baryons to the
mass density ($\Omega$) of the universe
is almost of the same order as the contribution of the dark matter to it ?
(ii) how does one understand the dark objects with mass $\sim 0.5
M_{\odot}$ observed in the MACHO experiment\cite{micro}, which are
supposed to constitute up to 20\% of the mass\cite{gyuk} of the halo of
the
Milky way galaxy and presumably be connected to the dark constituent that 
contributes to $\Omega$ ? and, finally (iii) what explains the density
profile of dark matter in galactic halos; in particular, the 
evidence that the core densities of
galactic halos remain constant as the radius goes to zero.

There are many particle physics candidates for dark constituent of
the universe. Generally speaking, the prime consideration that leads to
such candidates is that they yield the right order of magnitude for the
relic density and mass necessary to get the desired $\Omega_{DM} \approx 
0.2-1$. This is, of course, the minimal criterion.
It requires that the annihilation cross section of the
particles must be in a very specific range correlated with their mass. The
most widely discussed candidates are
the lightest supersymmetric particle (LSP) and the
Peccei-Quinn particle, the
axion. The first one is expected to have a mass in the range of 100 GeV
whereas the mass of the second would be in the range of $\sim 10^{-6}$ eV.
The present consensus seems to be that value of $\Omega_{CDM}$ is around
$0.2-0.3$, with $\Omega_{\Lambda}$ making up the rest of the energy
density of the universe at the moment.
Compare these values with $\Omega_{B}\simeq 0.05$. The CDM contribution
and the baryon contribution to $\Omega$ are roughly of same order. On the
other hand, the nucleon mass is very different from the masses of either
the axion or the SUSY LSP. So to understand within the SUSY or the axion  
models why $\Omega_B \sim
\Omega_{DM}$, one needs to work in a special range for the parameters
of the theory. In either of these pictures, the MACHO observations must
have a separate explanation. 

Furthermore, in recent years it has been emphasized that the LSP and the
axion may also have
difficulty in explaining the observed core density behaviour of dwarf
speroidal galaxies which are known to be dark matter dominated. The point
here is that both the axions and neutralinos, being collisionless and
nonrelativistic, accumulate at the core of galactic halos, leading to
a core density $\rho (R)$ which goes like $R^{-2}$ rather than a constant
which seems to fit data better\cite{primack,moore,mcgough,frenk}. We will
refer to this as the core density puzzle\cite{ss,ha,ho}.

 Spergel and Steinhardt\cite{ss} have recently revived
an old idea\cite{hall}, that dark matter may be strongly self interacting,
to resolve this puzzle. They argue that,
for the right range of the parameters of the particle, it may
lead to a halo core which is much less dense and hence in better
agreement 
with observations. Specifically they note that 
if the dark matter particle has mean free collision
path of about a kpc to a Mpc, then the core on this scale
cannot ``keep on accumulating'' dark matter particles, since these will
now scatter
and ``diffuse out''. For typical dark matter particle densities of order 
of one particle per cm$^3$, this requires a cross section for scattering
of $\sigma\simeq 10^{-21}-10^{-24}$ cm$^2$. Furthermore the properties of
the dark matter particle must be such that it must not allow for
dissipation of the thermal energy via emission of light particles;
otherwise,the galactic core would cool and lead to an increase in
core density. If these
considerations stand the test of time, a theoretical challenge would be to
look for alternative dark matter candidates (different from the popular 
ones described above) and the associated scenarios for physics beyond the
standard model. 

A class of models known as mirror universe models have recently been
discussed. These are motivated theoretically by string theory and
experimentally by neutrino physics. They predict the existence of a
mirror sector of
the universe with matter and force content identical to the familiar
sector (prior to symmetry
breaking)\cite{lee,berezhiani,volkas,sila}. Symmetry breaking might either
keep
the mirror symmetry exact or break it. This leads to two classes of
mirror models: the symmetric mirror model, where all masses and forces in
the two sectors remain the same after symmetry breaking\cite{volkas} and
the asymmetric mirror model\cite{berezhiani} where the masses in the
mirror sector are larger than those in the familiar sector. The
mirror particles interact with the mirror photon and not the familiar 
photon so
that they remain dark to our observations. Since the the lightest
particles of the mirror sector (other than the neutrinos), the mirror
proton and the mirror electron
(like those in the familiar sector) are stable and will have abundances
similar 
to the
familiar protons and electrons, the proton being heavier could certainly
qualify as a dark matter candidate. We will show in the next section they
can indeed play this role.

It has 
been shown\cite{tm,blin} that, consistent with the cosmological
constraints
of the mirror universe theory, the mirror baryons have the desired relic
density  to play the role of dark matter of the universe. 
 The additional neutrinos of the
mirror sector are the sterile neutrinos that appear to be needed in order
to have a simultaneous understanding of the three different observed 
neutrino oscillations i.e. solar, atmospheric and the LSND observations.
In fact, one view of neutrino oscillation explanations of these phenomena
fixes the ratio of familiar particle mass to mirror particle mass
thereby narrowing down the freedom in mirror sector parameters. If
indeed sterile neutrinos turn out to be  required, mirror universe theory
is one of the few models where they appear naturally with masses in the
desired range. If
we denote the mass ratio $m_{p'}/m_p~=~\zeta$, then a value of $\zeta\sim
10-30$ is required to explain the neutrino puzzles.
What is more interesting is that for the
same parameter range required to solve the neutrino puzzles,
mirror matter can also provide an explanation of the
microlensing observations\cite{tm}- in particular  why the observed MACHOs 
have a mass very near the solar mass and are still dark.

\section{Asymmetric mirror model in brief}

Let us start with a brief overview of the mirror matter models 
\cite{berezhiani,volkas}. The basic idea of the model is extremely simple:
{\it  duplicate the standard model or any extension of it in the gauge
symmetric Lagrangian and allow for the possibility that symmetry breaking 
may be different in the two sectors.} (See table below)  
There is an exact mirror symmetry connecting the Lagrangians (prior to
symmetry breaking) 
describing physics in each sector. Clearly the $W's, \gamma's $ etc in 
each sector differ from those in the other as do the quarks and 
leptons. When the symmetry breaking scale is
different in the two sectors, we will call this the asymmetric mirror
model\cite{berezhiani}. The QCD scale being an independent scale in the
theory could be arbitrary. We will allow both
the weak scale and the QCD scale of the mirror sector to 
differ from those of the familiar sector\cite{tm} and assume the same
common 
ratio $\zeta$ for both scales i.e. $<H'>/<H>=\Lambda'/\Lambda\equiv
\zeta$.

\begin{center}
\begin{tabular}{ccc}
$u,~d,~e,~\nu_e$ & $\leftrightarrow$  & $u',~d',~e',~\nu'_e$ \\
      &    &    \\
$W,~ Z,~ \gamma,~G$ & $\leftrightarrow$ & $W',~Z',~\gamma',~G'$ \\
      &    &    \\
$\phi, \nu_R,...$ & $\leftrightarrow$ &
$\phi',~\nu'_R,...$\\
       &        &   \\
       &        &    \\
       &  $\leftarrow${\bf Gravity}$\rightarrow$ & \\
\end{tabular}
\end{center}

It is assumed that the two sectors in the 
universe are connected by only gravitational interactions. It was shown in
\cite{berezhiani,volkas} that gravity induces nonrenormalizable operators
that generate mixings between the familiar and the mirror neutrinos. This
is one of the ingredients in the resolution of neutrino puzzles. To get an
idea of how this works, note that the lepton operators induced due to
nonperturbative gravitational effects have the form $\frac{LH
LH}{M_{Pl}}$, $\frac{LHL'H'}{M_{Pl}}$ and $\frac{L'H'L'H'}{M_{Pl}}$. After
spontaneous breakdown (i.e. $<H>= v$ and $<H'>=v'$, we get for the mass
matrix mixing the first generation neutrinos from each sector to have the
form (in the basis $\nu_e,\nu'_e$):
\begin{eqnarray}
M=\frac{v^2}{M_{Pl}}\left(\begin{array}{cc} 1 & \zeta \\ \zeta &
\lambda\zeta^2\end{array}\right)
\end{eqnarray}
where $\zeta \equiv v'/v$ defined above. To solve the solar
neutrino puzzle via small angle MSW solution, we will choose $\lambda\sim
1$ and $\zeta \approx 10-30$. This gives the sterile neutrino mass of
order $10^{-3}$ eV or so, choosing $M_{Pl} \simeq 10^{18}$ GeV, which is
also in the right range to explain the solar neutrino puzzle. Note that
these are meant to indicate that the model leads to numbers in the right
ball park. Emboldened by this result, we will consider the asymmetric
version and look at its cosmological implications.

In discussing cosmology, we first note that
 both sectors of the universe will evolve according to
the rules of the usual big bang model except that the cosmic soups in 
the two sectors may have different temperature. In fact the constraints of
big bang nucleosynthesis require that the post inflation reheat
temperature  
in the mirror sector $T'_R$ be slightly lower than that in the familiar
sector $T_R$ (define $\beta\equiv T'_R/T_R$) so that the contribution of
the light mirror particles such as $\nu', \gamma'$ etc. to nucleosynthesis
is not too important. This is called asymmetric inflation and can be
implemented in different ways\cite{asym}. We will see that, if we want the
mirror nucleons to play the role of the dark matter, we will need a
definite value of $\beta$ depening on the choice of $\zeta$. This in turn
will help us to predict a value for the equivalent extra neutrinos at the
BBN epoch (i.e. $\delta N_{\nu}$).

Before detailed discussion, let us first note the impact
of the asymmetry on physical parameters and processes. First it implies
that $m_i\rightarrow\zeta\,m_i$ with $i=n,p,e,W,Z$. This has important
implications for the nuclear and atomic physics of the mirror sector
\cite{berezhiani,teplitz}.
For instance, the binding energy of mirror hydrogen is $\zeta$ times
larger so that the recombination in the mirror sector takes place
much earlier than in the visible sector. With $\beta\equiv T'_R/T_R$ as
above, the mirror recombination occurs when the temperature of the
familiar sector is $\zeta/\beta T_r$ where $T_r$ is the recombination
temperature in the familiar sector. The mirror sector recombination takes
place before familiar sector recombination; this means that density
inhomogeneities in the mirror sector begin to grow earlier and familiar
matter can fall into them later as in typical cold dark matter scenarios.

\section{Mirror nucleon as dark matter}

 One can also compute the
contribution of mirror baryons to the mass density of the universe as
follows: 
\begin{eqnarray}
\frac{\Omega_{B'}}{\Omega_{B}}\simeq \beta^3 \zeta 
\label{cdm}
\end{eqnarray}
Here we have assumed that baryon to photon ratio in the familiar and the
mirror sectors are the same as would be expected since the dynamics are
same in both sectors due to mirror symmetry.
Eq. (2) implies that both the baryonic and the mirror baryon contribution
to $\Omega$ are roughly of the same order, as observed. This provides a
resolution of the first conceptual puzzle. Furthermore if
we take $\Omega_B\simeq 0.05$, then $\Omega_{B'}\simeq 0.2$ would require
that $\beta =(4/\zeta)^{1/3}$. From this one can calculate the effective
$\delta N_{\nu}$ using the following formula:
\begin{eqnarray}
\delta N_{\nu}= 3 \beta^4 + \frac{4}{7} \beta^4 (\frac{11}{4})^{4/3}
\end{eqnarray}
where the last factor $(11/4)^{4/3}$ is due to the reheating of the mirror
photon gas subsequent to mirror $e^{+'}e^{-'}$ annihilation. For $\zeta =
20$, this implies $\delta N_{\nu} \simeq 0.6$ and it scales with $\zeta$
as $\zeta^{-4/3}$. Thus in principle the idea
that mirror baryons are dark matter could be tested by more accurate
measurements of primordial He$^4$, Deuterium and Li$^7$ abundances which 
determine $\delta N_{\nu}$.

Clearly to satisfy the inflationary constraint of $\Omega_{TOT}=1$, we
need
$\Omega_{\Lambda}\simeq 0.7$. These kinds of numbers for cold dark matter
density apparently emerge from current type I supernovae
observations. It is interesting to note that if one were to require that
$\Omega_{CDM} = 1$, the mirror model would require that $\zeta$ be much
larger (more than 100) which would then create difficulties in
understanding both the neutrino data and the microlensing
anomalies. Thus mirror baryons seem to have just the right properties to 
 be the cold dark matter of the universe.

\section{Explaining the microlensing anomaly}

Let us now turn to show how the mirror model accounts for the microlensing
observations. We start with the four equations of stellar structure:

\be\label{st1}
dP/dr=-G\rho(r)M(r)/r^2
\ee
\be\label{st2}
dM(r)/dr=4\pi\,r^2\rho(r)
\ee
\be\label{st3}
L(r)/4\pi\,r^2=-(16/3)\sigma_{SB}(T^3/\rho\kappa)dT/dr
\ee
\be\label{st4}
dL/dr=4\pi\,r^2\epsilon(r)\rho(r)
\ee
where $\kappa(r)$ is the opacity (cross section per unit mass) at radius
$r$,
$\sigma_{SB}$ the Stefan-Boltzmann constant, L(r) the luminosity at
radius $r$,
and $\epsilon(r)$, the rate of energy generation per unit mass at radius
$r$.
We will need three terms in the equation of state (below) taken one or two
 at a time:
\be\label{eos}
P=(\rho/m)kT + (4\sigma_{SB}/3c)T^4 + (h^2/2m_e)(3/8\pi)^2(\rho/m)^{5/3}
\ee

where the three terms represent gas pressure, radiation pressure, and
(non-relativistic) degenerate electron pressure.  $m$ is the nucleon mass,
$m_e$ that of the electron.  We have neglected such niceties
as keeping track of how many objects there are for each $m$ of gas (2 for
 H,
3/4 for He, etc)

We will make standard, illuminating if crude, approximations
\cite{clayton}
 in order to understand the $\zeta$ behavior of the
solutions to the above equations.  First we write
\be\label{peq}
P=\rho\,GM/R,\ \ \ \rho=3M/4\pi\,R^3
\ee
where $P$ and $\rho$ are roughly core averages. Here $M$ and $R$ are mass
and radius of the star.This gives the useful relation

\be\label{prho}
P=(4\pi/3)^{1/3}GM^{2/3}\rho^{4/3}
\ee
To find the maximum mass of a (main sequence) mirror star, which is of
interest to us, we note that as the mass of the star
gets bigger, the core temperature rises. Therefore, of the three
terms in the expression for the pressure in the Equation \ref{eos},
we expect $P_g$ and $P_r$ to dominate. Following Phillips\cite{clayton},
we parameterize them as fractions of the total pressure $P$ as below:
\be\label{beta}
P_g=\beta\,P,\ \ \ P_r=(1-\beta)P
\ee
We eliminate $T$ and solve for P from this parameterization to obtain
\be\label{pbet}
\beta P=[(\rho\,k/m)^4(\beta^{-1}-1)/(4\sigma_{SB}/3c)]^{1/3}
\ee
Using Equation (\ref{prho}) again then gives
\be\label{mmax}
M_{max}\sim[(1-\beta)c/\sigma_{SB}]^{1/2}G^{-3/2}(k/m)^2/\beta^2
\ee
As $\beta$ approaches $1$, the energy density is increasingly dominated by
photons (relativistic particles) and stars become unstable.  Taking a
cutoff around $\beta\sim1/2$ gives a maximum stellar mass around
$70M_{\odot}$. Thus
the range for stars is roughly $0.07M_{\odot}$ to $70M_{\odot}$.  From
Equation
(\ref{mmax}) one sees, in the approximation that instability sets in at
the
same $\beta$ independent of $\zeta$, that $M_{max}$ varies as
$\zeta^{-2}$. For $\zeta=15$, we get for the maximum mass of the mirror
star $0.5 M_{\odot}$, which is of the same order of mass as the MACHO
microlensing events. Our model therefore provides a
resolution of the microlensing anomaly that avoids the strong constraints
of Freese et al\cite{freese} for familiar sector white dwarfs.

We want to point out here that we do not expect all of mirror dark matter
to condense to form mirror stars. Instead, we would expect it to be
in the form
of a mixture of mirror dust and mirror stars. In this connection, it has
been noted recently\cite{miller} that current upper limits to scattering
optical depths for Thomson scattering in early universe suggests that 
compact objects of any kind cannot be the main dark matter constituent.
This would also suggest a mixed picture of the kind mentioned.

\section{Self interaction of mirror matter and halo core density
problem}

As noted in the introduction, there appear to be indications from the core
density profile of dwarf and low surface brightness
galaxies\cite{moore,mcgough} that the dark matter may need to
be endowed with a significant self interaction. According to the analysis
of Spergel and Stenhardt\cite{ss}, the self interaction must be such that
the collision cross section of dark matter particles should be of order
$10^{-23}(m/GeV)$ cm$^2$ for a one GeV particle corresponding to a DM mean
free
path of order of $\lambda_{ss}= 300$ kpc in a gas with number density
given by
$\rho_{DM}/m$. Also we note that the cross section would scale
linearly with the mass of the dark matter particle. The
mean free path requirement cannot be met by the neutralino or the axion
but is met quite naturally by the mirror dark matter forces\cite{tm2}.

Two obvious kinds of self interactions are the self interactions
due to inter-atomic forces and nuclear forces. The latter are not very
effective as can be seen by the following crude argument. From pure
$\zeta$ scaling, we can infer that $\sigma_{N'N'}\approx
\sigma_{NN}\zeta^{-2}$. The value of $\zeta =20$ would put the
mirror nucleon cross section to be of order $10^{-26}$ cm$^2$, which is
much
too small. Note that for $\zeta=20$, one would need a cross section of
order $10^{-22}$ cm$^2$ to get $\lambda = \lambda_{ss}$. On the other
hand,
one would expect that there
can be scattering of the dark matter particles  (mirror atoms) due to
inter-atomic forces.
A crude estimate of such cross sections is given by $\sigma_{H'H'} \approx
\pi~ (a'_0)^2$, where $a'_0$ is the Bohr radius of the mirror hydrogen
atom. One can then estimate the mean free path of the dark matter
particles in the mirror model to be $\lambda_{DM}\approx 3 \pi \zeta^3
\times 10^{16}$ cm. For $\zeta =20$, this gives $\lambda_{DM} \approx 0.3$
kpc, which is not far below the required values for explaining the halo
core density profile. Note that Spergel and Steinhardt \cite{ss} require
$\lambda_{DM}\approx $ kpc to mpc. 
 On the other hand, one could take the value of $\zeta=30$ and
thereby get $\lambda \approx $ one kpc.

We thus feel that mirror dark matter presents the best scenario for
understanding the halo core density profile using a self interacting
dark matter model. Of course more detailed numerical work is needed 
to confirm these qualitative conclusions.

\section{Structure Formation}

Finally, we use the remainder of this article
 to make plausible that, in spite of the
$\zeta^2$ decrease in cross sections for most processes, the facts that
(a) structure
formation begins earlier for the mirror sector (because recombination
occurs before matter-radiation equality) and (b) the higher mirror
temperatures for the same processes, than familiar
temperatures, permit formation of galactic and smaller structures.  In
doing this, we will make use of our previous work in \cite{teplitz} and
\cite{tm}, as well as that of Tegmark et al \cite{teg}.

Much of the work of \cite{teplitz} can be carried over to the present
work,
after suitable modification to take into account the fact that, in the
current model, the proton mass scales as $\zeta$.  Here, we will assume
that primordial perturbations are "curvature" or "adiabatic"
perturbations.  This means that the scale of the largest structures are
set by mirror Silk damping \cite{silk}.  $\gamma'$ diffusion wipes out
inhomogeneities until the $\gamma'$ mean free path,
\begin{equation}
                \lambda'=[\sigma_T\zeta^{-2}n_e^{'}]^{-1}
\end{equation}

where $\zeta^{-2}\sigma_T$ is the mirror matter Thomson cross section and
$n_e'$ is its electron number density, becomes greater than one third the
horizon distance ($ct$).  Silk damping turns off because the $\lambda'$
increases as $z^{-3}$ while $ct$ only increases as $z^{-2}$.

First, we compute, from Silk damping, the masses of the largest
structures in this picture.  Structure formation starts with 
mirror sector particles, and familiar sector particles
later fall into these.  For numerical
values below, we will take, $h=0.7$ and $\Omega_{{B'}}=0.2$.
We pick $t\sim(z_1/z)^2 s$ with $z_1=4\times10^9$ and
$n_{{B'}}=\Omega_{{B'}}\rho_{c}z^3/(\zeta m_p)$ with
$\rho_c=1.9h^2\times10^{-29} g/cm^3$.  
Silk damping stops at around $z_{sd}\sim 8\zeta^3$ which gives
\begin{eqnarray}
        \lambda_{sd}\sim2.5\times10^{27}/\zeta^6 ~~cm\\ \nonumber
        M_{sd}\sim10^{54}/\zeta^9~~gm
\end{eqnarray}

Note that, for $\zeta\sim 10$, this is about the mass (and size) of a
large
galaxy.  This coincidence could be an important factor in
understanding galaxy sizes should this model correspond to reality.

As in \cite{teplitz}, we parametrize the separation of $M_{sd}$ from the
expanding universe as taking place at

\begin{equation} 
        z_{stop}=z_Mz_{sd}
\end{equation}

with

\begin{equation} 
        R_G=\lambda_{sd}/z_M
\end{equation}

After violent relaxation we have for the proton temperature

\begin{equation}
        T_p=GM_G\zeta m_p/R_G \sim 10^{-4}z_M/\zeta^2 ~~ergs
\end{equation}

with $\rho$, outside the central plateau, given by
\begin{equation}
        \rho(R)=A/R^2 \sim 10^{26}z_M/(\zeta^3R^2)~~gm/cm^3
\end{equation}

We now turn to the question of whether this isothermal sphere is likely to
fragment and form mirror stars.  For this we compute the amount of mirror
molecular hydrogen since it is its collisional excitation (and
subsequent radiation) that is believed
to be the chief mechanism that provides cooling for formation of stars.
If the rate for this mechanism is faster than the rate for free fall into
the mass of the structure at issue, we can expect local regions to cool
fast enough to result in fragmentation of that
structure.  We do here a very rough estimate of mirror galaxy
fragmentation into mirror globular clusters, using the results of
\cite{teg}, but leave to a more detailed work further fragmentation into
the $0.5M_{\odot}$ structures predicted in \cite{tm}.

Reference \cite{teg} give a useful approximation to their numerical
results for the fraction of molecular hydrogen, $f_2$ ($f_0$ denotes its
primordial value):
\begin{equation}
        f_2(t)=f_0+(k_m/k_1)ln[1+x_0nk_1t)
\end{equation}
where, as a first try, $k_m$ can be taken as just the rate for
$H+e^-\rightarrow H^-+\gamma$ (which they conveniently give as about
$2\times10^{-18}T^{.88} cm^3s^{-1}$), while $k_1$ is the rate for
$H^++e^-\rightarrow H+\gamma$ ($2\times10^{-10}T^{-0.64} cm^3s^{-1}$).
Equation
(8) is the result of $H_2$ production from the catalytic
reactions $H+e^-\rightarrow H^-+\gamma$ followed by $H+H^-\rightarrow
H_2+e^-$ competing
against the recombination reaction that destroys the catalyst, free
electrons,
(approximately) as $1/t$ (assuming constant density).  Our goal here is to
show from Equation (20) that it is plausible that $f_2$, the
fraction of molecular hydrogen, rises from its primordial value of
$10^{-6}$ to the region above $10^{-4}$ where cooling tends to be
competitive with free fall.

First, we note that, if $k=<v\sigma>\sim AT^{\gamma} cm^3/s$, for familiar
e and p, we expect that, for mirror e and p, scaling with $\zeta$ to go as
$\zeta^{-{(2+\gamma)}}AT^{\gamma}$, since $\sigma$ must go as
$\zeta^{-2}$,
all factors of $T$ must be divided by some combination of $m_e$ and
$M_p$, both of which go as $\zeta$ (making this model much easier to
compute from than that of \cite{teplitz}).

We now estimate fragmentation.  From Equation (18) we see that the
galactic temperature should begin at about 10 eV at a time when the cosmic
temperature is about 1 eV and the cosmic gamma number density is about
$10^9/cm^3$.  The rate for ``compton cooling" is very fast at this high
density (unlike at later times for the familiar case) and there should be
rapid cooling to about 1 eV.  We can now compute the Jeans length for
fragments as a function of distance R from galaxy central.  We use

\begin{equation}
        \rho_J=(T/Gm)^3/M^2
\end{equation}

If we set the Jeans mass, $M$, to $4\pi r^3\rho_J/3$, we can solve for $r$
obtaining (if we are careful to convert $T$ in Equation (6)
from ergs)

\begin{equation}
        r=R[10^{-7}\zeta^2/z_M]^{1/2}\sim10^{-2}R
\end{equation}

Now inserting into Equation (20) gives the coefficient of the
log term on the order of $10^{-3.5}$ and the coefficient of $t$ in 
the argument varying from $10^{-13}$
to $10^{-17}$ as $R$ varies from 1 to 100 kiloparsecs while the free fall
time ($(G\rho)^{-1/2}$) varies from $10^{14}$ to $10^{16}$.  This would
appear to indicate the likelihood of fragmentation of the original Silk
damping structure into smaller units, and the eventual formation of the
$0.5M_{odot}$ black holes that explain the microlensing events of
\cite{tm}.

\section{Conclusion}

We have argued that the asymmetric mirror model\cite{berezhiani},
originally proposed to solve neutrino puzzles and subsequently
advocated\cite{tm,blin}  as providing an alternative dark matter candidate
has the advantage of resolving the microlensing anomaly and possibly
the core density problem of dark halos. Possible tests of these models
are to narrow the allowed values of $\delta N_{\nu}$ from more accurate
observation of deuterium, He$^4$ and Li$^7$ and observing whether further
accumulation of MACHO candidates lie in the mass between 0.1- 1 solar
mass. Needless to say that if the underground searches for the cold dark
matter now under way lead to a positive signal, mirror matter cannot be
the dominant component of the dark matter of the universe.

 The
work of R. N. M. is supported by the National Science Foundation grant
under no. PHY-9802551 and the work of V. L. T. is supported by the DOE
under grant no. DE-FG03-95ER40908.


\begin{thebibliography}{99}

\bibitem{book} For a survey of recent observational and theoretical
advances
in this area see, {\it Dark Matter in Astro and Particle Physics} ed. by
H. V. Klapdor-Kleingrothaus and Y. Ramachers (World Scientific), 1996
and H. V. Klapdor-Kleingrothaus and L. Baudis, (IOP Publication), 1998.

\bibitem{micro} C. Alcock et al. Ap. J. {\bf 486}, 697 (1997); R. Ansari
et al. A \& A {\bf 314}, 94 (1996).

\bibitem{gyuk} E. Gates, G. Gyuk and M. Turner, Phys. Rev. {\bf D 53},
4138 (1996).

\bibitem{primack} R. Flores and J. Primack, Ap. J. {\bf 427}, L1 (1994).

\bibitem{moore} B. Moore, Nature, {\bf 370}, 629 (1994); B. Moore {\it et
al}, astro-ph/9907411.

\bibitem{mcgough} W. J. G. de Blok and S. S. McGough, MNRAS, {\bf 290},
533 (1997).

\bibitem{frenk} J. F. Navarro, C. S. Frenk and S. White, Ap. J. {\bf 462},
563 (1996).

\bibitem{ss} D. Spergel and P. Steinhardt, astro-ph/9909386.

\bibitem{ha} S. Hannestad, astro-ph/9912558; S. Hannestad and R. Sherrer,
astro-ph/0003046.


\bibitem{ho} Craig J. Hogan and J. J. Dalcanton, astro-ph/0002330.

\bibitem{hall} E. Carlson, M. Machacek and L. J. Hall, Ap. J. {\bf 398},
43 (1992); A. A. de Laix, R. K. Shaeffer and R. J. Sherrer, Ap. J. {\bf
452}, 495 (1995).

\bibitem{lee} T. D. Lee and C. N. Yang, Phys. Rev. {\bf 104}, 254 (1956);
K. Nishijima, private communication; Y. Kobzarev, L. Okun and I. Ya
Pomeranchuk, Yad. Fiz. {\bf 3}, 1154 (1966);  M. Pavsic, Int. J. T. P.
{\bf 9}, 229 (1974); S. I. Blinnikov and M. Y. Khlopov, Astro. Zh. {\bf
60}, 632 (1983); E. W. Kolb, D. Seckel and M. Turner, Nature, {\bf 514},
415 (1985); R. Foot, H. Lew and R. Volkas, Phys. Lett. {\bf B 272}, 67
(1991).

\bibitem{berezhiani} Z. Berezhiani and R. N. Mohapatra, Phys. Rev. {\bf
D 52}, 6607 (1995); Z. Berezhiani, A. Dolgov and R. N. Mohapatra, Phys.
Lett. {\bf B 375}, 26 (1996).

\bibitem{volkas} R. Foot and R. Volkas, Phys. Rev. {\bf D 52}, 6595
(1995).

\bibitem{sila} For other papers on the subject, see H. Hodges, Phys. Rev.
{\bf d 47}, 456 (1993); M. Collie and R. Foot, Phys. Lett. {\bf B432},
134 (1998); B. Brahmachari and R. N. Mohapatra, hep-ph/9805429; Z.
Silagadze, hep-ph/9908208; N. F. Bell and
R. Volkas, Phys. Rev. {\bf D 59}, 107301 (1999); N. Bell, hep-ph/0003072; 
R. Foot and S. N. Ginnenko, hep-ph/0003278; for a recent review, see Z.
Silagadze, hep-ph/0002255.

\bibitem{tm}  R. N. Mohapatra and V. L. Teplitz, Phys. Lett. {\bf B 462},
302 (1999).

\bibitem{blin} S. I. Blinnikov, astro-ph/9801015; Z. Berezhiani,
hep-ph/9602326.

\bibitem{asym} Z. Berezhiani et al. ref.\cite{berezhiani}; V. Berezinsky
and A. Vilenkin, hep-ph/9908257.


\bibitem{clayton} D. Clayton, {\it Principle of Stellar Evolution and
Nucleosynthesis}, McGraw-Hill, New York (1968);  A. C. Phillips, {\it
Physics of Stars}, John-Wiley (1996); W. K. Rose, {\it Advanced Stellar
Astrophysics}, Cambridge University Press (1998).

\bibitem{freese} D. S. Graff and K. Freese, Ap. J. {\bf 456}, L49
(1996); D. S. Graff, G. Laughlin and K. Freese, Ap. J. {\bf 499}, 7
(1998); For a review, see K. Freese, B. Fields and D. Graff,
astro-ph/9901178.

\bibitem{miller} M. C. Miller, astro-ph/0003176.

\bibitem{tm2} R. N. Mohapatra and V. L. Teplitz, astro-ph/0001362.

\bibitem{teplitz} R. N. Mohapatra and V. L. Teplitz, Ap. J. {\bf 478}, 29
(1997).

\bibitem{teg} M. Tegmark et al. Ap. J. {\bf 474}, 1, (1997).

\bibitem{silk} J. Silk, Ap. J. {\bf 211}, 638 (1977).
\end{thebibliography}
\end{document}